\documentclass[twocolumn]{aastex61}

\newcommand\prep{in preparation}

\shorttitle{New stellar cluster candidates in the Magellanic System}
\shortauthors{Andr\'es E. Piatti}

\begin{document}

\title{Stellar cluster candidates discovered in the Magellanic System}

\author{Andr\'es E. Piatti}
\affiliation{Consejo Nacional de Investigaciones Cient\'{\i}ficas y T\'ecnicas, Av. Rivadavia 1917, 
C1033AAJ, Buenos Aires, Argentina}
\affiliation{Observatorio Astron\'omico, Universidad Nacional de C\'ordoba, Laprida 854, 5000, 
C\'ordoba, Argentina}
\email{e-mail: andres@oac.unc.edu.ar}

\begin{abstract}

We address the presently exciting issue of the presence 
of stellar clusters in the periphery of the Magellanic Clouds (MCs) and beyond
by making use of a wealth of wide-field high-quality images
released in advance from the Magellanic Stellar Hystory (SMASH)
survey. We conducted a sound search for new stellar cluster candidates
from suitable kernel
density estimators running for appropriate ranges of radii and stellar densities.
In addition, we used a functional relationship to account for the
completeness of the SMASH field sample analyzed that takes into account not
only the number of fields used but also their particular spatial distribution;
the present sample statistically represents $\sim$ 50$\%$ of the whole SMASH survey.
The relative small number of new stellar cluster candidates identified,
most of them distributed in the outer regions of the Magellanic Clouds,
might suggest that the lack of detection of a larger number of new cluster candidates 
beyond the main bodies of the Magellanic Clouds could
likely be the outcome once the survey be completed.
\end{abstract}


\section{Introduction}

The Survey of Magellanic Stellar History \citep[SMASH,][]{nideveretal2015}
is aimed at mapping the expected stellar debris and extended populations from
interactions of the Magellanic System (MS) and the Milky Way (MW)
with unprecedented fidelity. As stellar clusters are considered, 
some recent outcomes have shown that there are still a substantial number 
of extreme low luminosity stellar clusters undetected in the wider MS periphery and the
MW halo \citep{kimetal2015a,pieresetal2016,martinetal2016}. Furthermore, streams
of gas and stars that might harbor stellar clusters have also been detected
\citep{mackeyetal2016,belokurovetal2016,deasonetal2016}.

As of October 2016, 58 SMASH fields (each one is an array of sixty-two 
2k$\times$4k CCD detectors covering $\sim$ 3 deg$^2$ with an unbinned pixel scale 
of 0.27 arcsec) have been made publicly available from the National Optical 
Astronomy Observatory (NOAO) Science Data Management (SDM) 
Archives\footnote{http //www.noao.edu/sdm/archives.php.}. Fig.~\ref{fig1}
shows the distribution of the complete list of SMASH fields and those
already observed and publicly available drawn with open and filled hexagons,
respectively. 

In this Letter we make use in advance of this wealth of released images
to anticipate some ultimate answers about the expected undetected stellar
clusters in the Magellanic Clouds (MCs) periphery. In Section 2 we describe our
search for new stellar cluster candidates, while Section 3 deals with the discussion
of the results. Finally, Section 4 summarizes the main conclusions of this
work.

\section{Data collection and analysis}

We downloaded calibrated, single-frame reduced images with instrument 
signature removed and WCS calibrations applied according to the
DECam Community Pipeline \citep{valdesetal2014}. We preferred those
reprojected images which have been corrected for distortion, etc, 
and are better astrometrically fixed. Finally, we filtered $g$ images
with exposure times longer than 267 sec; the
deepest images obtained by the DECam 
\citep{martinetal2016,drlica-wagneretal2016,kimetal2016}.
Then we run the {\sc DAOFIND} routine within the {\sc DAOPHOT}
suite of programs \citep{setal90} to produce photometric catalogs
of each $\sim$ 2.2 deg FOV image down to $g$ $\sim$ 24.0 mag.
Note that SMASH fields do not overlap those from the Dark Energy Survey 
\citep[DES,][]{des2016} and are deeper than the single $g$ 90 sec 
exposures of the Magellanic SatelLites Survey fields.
\citep[MagLiteS,][]{drlica-wagneretal2016}.

\begin{figure}
\includegraphics[width=\columnwidth]{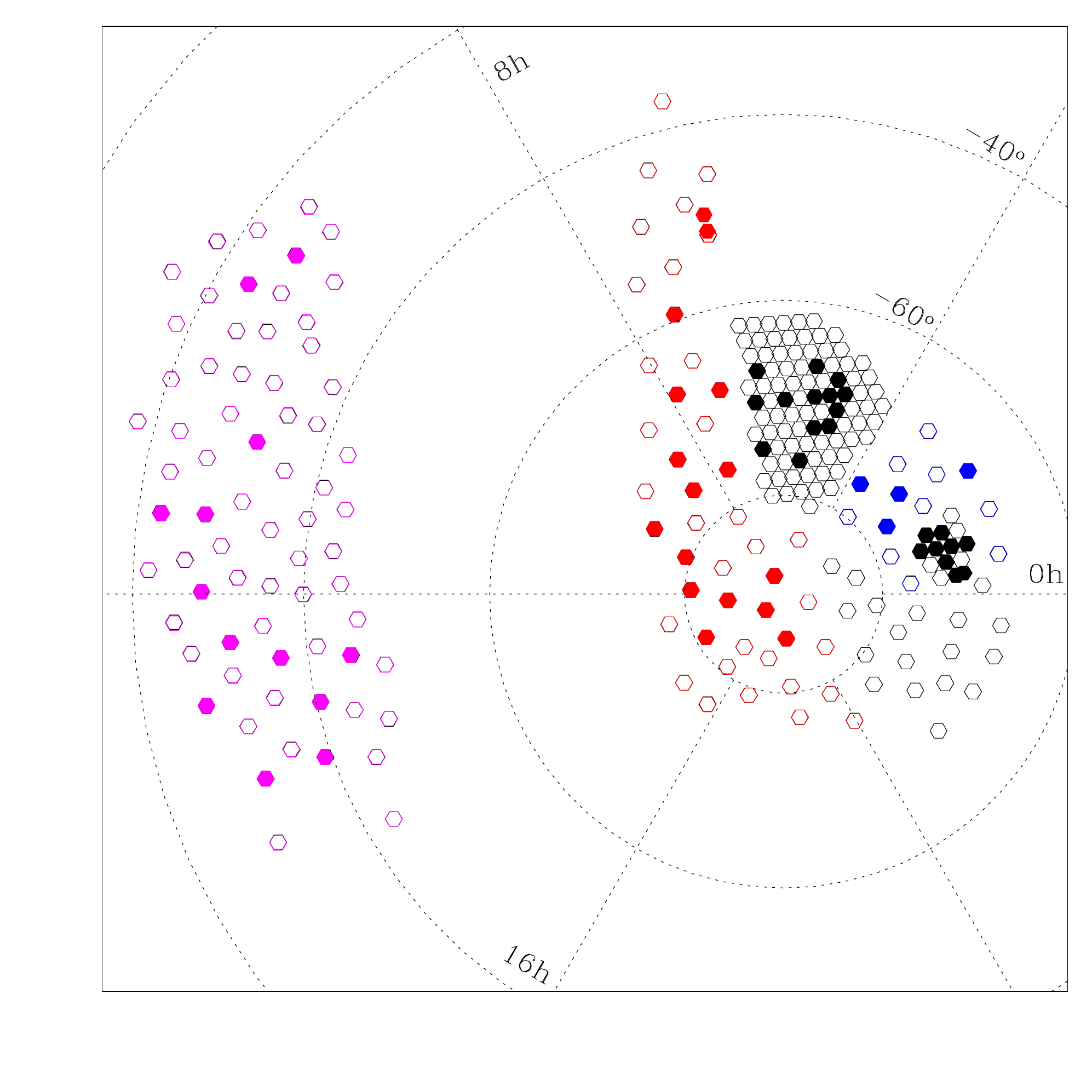}
\caption{The equal-area Hammer projection of the SMASH fields
in Equatorial coordinates. Filled hexagons
represent those analyzed in this work.}
\label{fig1}
\end{figure}

The strategy for searching new stellar clusters in the MS was developed in 
\citet{piattietal2016} with the
aim of applying it to the whole VISTA\footnote{Visible and Infrared Survey Telescope 
for Astronomy} near-infrared $YJK_s$ survey of the Magellanic Clouds system 
\cite[VMC,][]{cetal11}, and also used elsewhere \citep[e.g.][]{piatti16}.
It executes a series of AstroML 
routines \citep[][and reference therein for a 
detail description of the complete AstroML package and user's 
Manual]{astroml}, a machine learning
and data mining for Astronomy package. The method makes use of
the range of radii and stellar densities of known Large and Small Malleganic Cloud 
(L/SMC) clusters, so that it is able to detect the smallest and/or less dense
clusters in the MS. \citet{piattietal2016} searched for new star clusters in a 
pilot field of $\sim$ 0.4 deg$^2$ in the South-west side of the SMC bar, where the 
star field is the densest and highest reddened region in the galaxy. 
Based on the selection criteria mentioned above, they identified the 68 known star 
clusters located in that pilot field \citep{betal08} and 38 new ones ($\approx$ a 55 per cent 
increase in the known star clusters located in the surveyed field).
We refer the reader to \citet{piattietal2016} for quantitative
details concerning detection efficiency, crowding effects, cluster mass and concentration,
etc. We used here two different kernel 
density estimators 
(KDEs), namely, {\it gaussian} and {\it tophat}, and bandwidths from
 0.2 up to 1.0 arcmin. 

\begin{figure}
\includegraphics[width=\columnwidth]{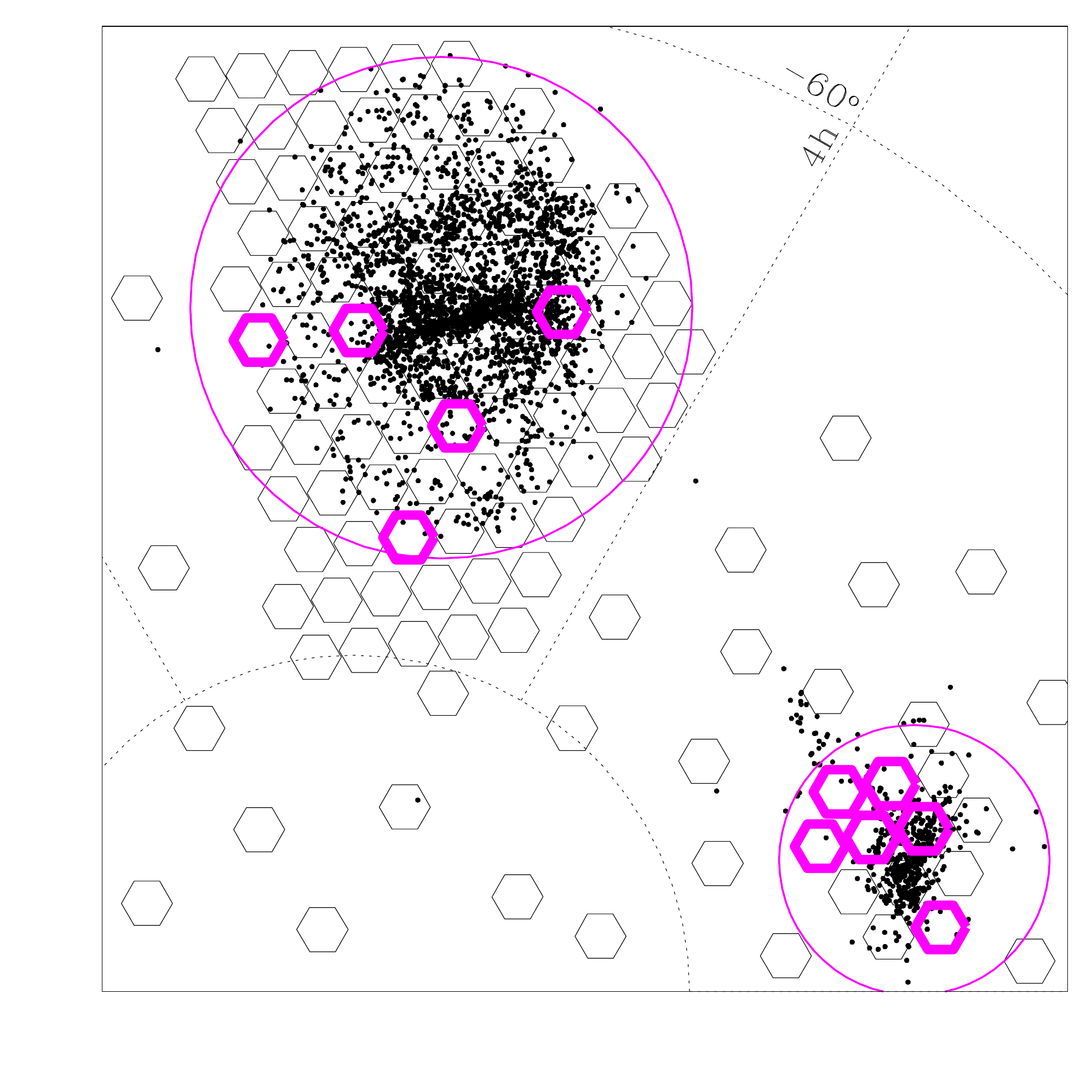}
\caption{Enlargement of Fig.~\ref{fig1} with
the positions of stellar clusters in the \citet{betal08}'s
catalogue superimposed for comparison purposes. The
SMASH fields where we identified new stellar cluster candidates
are drawn with thick magenta open hexagons.}
\label{fig2}
\end{figure}

From the total number of stellar overdensities detected per field, 
we merged the resulting lists, avoiding repeated 
findings from different runs with different bandwidths. 
We finally identified  some
new stellar cluster candidates in some few SMASH fields depicted
in Fig.~\ref{fig2} with thick magenta open hexagons,  where we also
recognized the 533 known cataloged stellar clusters. 
From these results we conclude that it is hardly possible that any stellar cluster 
down to the known smallest and/or less dense limits of MC clusters
has not been detected in the analyzed SMASH fields.

The coordinates of the new stellar cluster candidates 
are listed in Table~\ref{table1}, while 
2$\times$2 arcmin$^2$ $g$ images centred on
them are shown in Fig.~\ref{fig1A} of the Appendix on the online version of the journal. 
As far as we are aware, these objects have not been included in previous peer-reviewed
stellar cluster cataloging works.
The estimation of their structural and
fundamental parameters will be presented 
in a forthcoming paper \citep{piattietal2017}.
As can be seen, all of them are mainly located in the outer regions of the L/SMC 
main bodies; no stellar cluster candidate was detected beyond those regions.  
In Fig.~\ref{fig2} we have included circles of 7.4 to 4.0 deg centered on the LMC and
SMC, respectively, to illustrate the areas over which the new
stellar cluster candidates are distributed.
This result is in agreement with the positions of recently new stellar clusters detected
towards the outermost northern LMC regions \citep{pieresetal2016,martinetal2016},
as well as stellar streams \citep[][and references therein]{mackeyetal2016,deasonetal2016},
in the sense that previously less explored outer regions of the L/SMC can harbor undetected stellar clusters.

The apparently small number of new stellar cluster candidates detected throughout 
58 SMASH fields ($\sim$ 174 deg$^2$) caught our attention, since mapping 
the MS to a surface brightness limit of $\sim$ 35 mag/arcsec$^2$ 
\citep[$g$ $\sim$ 24.0 mag][]{nideveretal2015}
should result in an unprecedented deep stellar cluster survey.
In order to evaluate the impact of such a small statistics of new stellar cluster
candidates, in the context of the ambitious goals raised by SMASH, 
we estimated the completeness factor of the SMASH fields analyzed.

\begin{deluxetable}{lcc}
\tablecaption{Positions of new MS stellar cluster candidates.}
\label{table1}
\tablehead{\colhead{ID$^a$} & \colhead{R.A.}  & \colhead{Dec.} \\
\colhead{}     & \colhead{(h m s) } & \colhead{($\degr$ $\arcmin$ $\arcsec$)} }

\startdata
Field\,4-01 &  0 33 01.051 & -72 59 22.21 \\
Field\,10-01 & 1 07 33.964 &-72 09 53.78  \\
Field\,10-02 & 1 12 23.670 &-72 17 31.40  \\
Field\,10-03 & 1 04 46.715 &-72 42 58.19 \\
Field\,11-01 & 1 03 55.763 &-72 56 00.47   \\
Field\,11-02 & 1 01 18.189 &-73 27 57.20  \\
Field\,11-03 & 1 04 11.120 &-74 17 53.45  \\
Field\,12-01 & 1 13 42.512 &-74 45 14.32  \\
Field\,15-01 & 1 23 00.795 &-73 13 07.21  \\
Field\,16-01 & 1 29 53.196 &-74 50 43.17   \\
Field\,16-02 & 1 29 50.857 &-74 40 55.26  \\
Field\,30-01 & 4 48 15.866 &-69 40 23.93  \\
Field\,30-02 & 4 52 58.700 &-69 22 53.10  \\
Field\,40-01 & 5 21 06.005 &-71 54 50.73  \\
Field\,40-02 & 5 25 26.887 &-72 38 21.06  \\
Field\,40-03 & 5 12 34.451 &-72 37 34.53  \\
Field\,40-04 & 5 17 56.371 &-73 33 24.64  \\
Field\,40-05 & 5 21 57.364 &-73 30 52.79   \\
Field\,40-06 & 5 12 21.198 &-73 16 15.43  \\
Field\,40-07 & 5 24 19.613 &-73 05 30.68  \\
Field\,44-01 & 5 30 41.792 &-75 40 07.77  \\
Field\,44-02 & 5 26 04.970 &-75 24 57.49  \\
Field\,51-01 & 5 49 13.971 &-70 40 25.65  \\
Field\,55-01 & 6 26 56.148 &-70 18 00.95  \\
\enddata

\noindent $^a$We have kept the SMASH field identifications
to which we have added a running number.
\end{deluxetable}

For analysis purposes we have distinguished three regions, called
R1, R2 and R3, which were painted in Fig.~\ref{fig1} with blue, 
red and magenta colors, respectively. They embrace 13, 62 and 69
SMASH fields of which 4, 18 and 13 were used here for searching for 
new stellar clusters. Other analyzed 23 fields placed in the L/SMC main bodies, 
are represented by black filled circles and were not considered in this
 estimate because we are interested in assessing our results
for the MC periphery. Note also that the western side of the SMC
has not been observed by SMASH yet, so that it was not taken into account
either.

The analyzed fields are distributed 
in such a way that they are not concentrated in a particular
zone of the three regions, but rather across areas larger than that 
for the SMASH 20\% filling factor. Because of this particular spatial distribution 
the completeness factor of the analyzed fields  depends on
both the number of fields analyzed and their distribution
within each region. The more the number of fields analyzed, the higher the
completeness, so that the completeness results proportional to the ratio between
analyzed fields and the complete list of SMASH fields. On the other hand, the 
analyzed fields in a particular R region can be placed all together (at the filling factor) 
or spread throughout
the whole R region, so that the completeness turns out higher as the area covered
by the analyzed fields is statistically more similar to that of the R region.

In order to estimate such a completeness factor we used the
expression:

\begin{equation}
\psi(k,p) = k / (1 - (p -k )^2) \hspace{1cm}  {p,k} \in \{0,1\};\\
 p \ge k,
\end{equation}

\noindent where $k$ represents the ratio between
filled and open hexagons (see Fig.~\ref{fig1}) and $p$ is a measure of the similarity 
between their spatial distributions, respectively.  Note that when the analyzed fields
are placed as close as possible, $p$ equals $k$, and when they spread throughout an R region,
equals 1. Thus, the better an R area is covered by the analyzed fields, the higher the $p$ value.
The expression ($p - k$ )$^2$ measures in quadrature the difference between the statistical
spatial coverages given by $p$ and that when all the analyzed fields are placed together ($k$),
so that 1 $-$ ($p - k$ )$^2$ aproaches the unity when $p$ is closer to $k$. Thus, for a fixed
ratio of analyzed fields $k$, the completeness increases as the distance in quadrature between $p$ and $k$
increases as well.
To evaluate $p$, we used the {\it kde.test} statistical
function provided by the {\it ks} package (version 1.10.4)\footnote{http //www.mvstat.net/tduong.}. 
The function applies a two-dimensional KDE
based algorithm, able to broadly asses the similarity between
data in two different arrays. The result,
quantified by a p-value, is the probability, assuming the null
hypothesis is true, of observing a result at least as extreme as 
the value of the test statistic \citep{fb2012}.
Here the null hypothesis is that  (x,y) coordinate samples for filled and
open hexagons come from the same underlying 
distribution, with a lower p-value indicative of a lower probability
that the null hypothesis is true. 

We obtained p-values of 0.98, 0.93
and 0.93, from which  $\psi$($k$,$p$) resulted to be 0.56, 0.53 and 
0.42 for R1, R2 and R3, respectively. The p-values show that 
the analyzed fields very well match the designed SMASH field distribution,
with a lower spatial frequency though. Therefore, we can conclude that,
it is not obvious that a substantial number of undetected stellar clusters,
compared to the number of new candidates identified here,
populate the MC periphery. On the contrary, we speculate with the
possibility that there could exist very few isolated stellar clusters in the
MC periphery stripped out from their parent galaxies due to Magellanic 
Clouds/MW interactions \citep[e.g.][]{hammeretal2015,salemetal2015}. This
speculation could fuel the present 
debate about whether ultra-faint objects discovered in the MW halo are 
dark matter free (stellar cluster) or dark matter dominated (dwarf galaxy)
objects \citep[e.g.,][]{kimetal2015a,martinetal2015,salesetal2016,drlica-wagneretal2016,contentaetal2016}. 

\section{conclusions}

We made use of publicly available SMASH images to conduct a sound search
for new stellar clusters in the MS. After building photometric catalogs
from stars found by {\sc DAOFIND} in each image,
we embarked in such a huge time consuming challenge by employing 
density kernel estimators with appropriate bandwidths in order to
 detect the smallest and/or less dense clusters in the MS.
We found out 24 new stellar cluster candidates distributed in 11 different
SMASH fields, most of them located in the outer regions of the L/SMC
discs. Although their spatial distribution confirms that outer regions
of the MC have been less explored in the past, their small number
suggests that there would appear to be low chances of detecting
significant number of stellar clusters there.

Bearing in mind that the analyzed sample of SMASH fields is far from
being complete, we estimated the completeness factor of them by
using a completeness function that depends on both the number of
fields analyzed and their spatial distribution. Because of such
a particular spatial coverage, we found that the present sample
-statistically speaking- nearly
represent 50$\%$ of the whole survey. This means that the lack
of detection of a larger number of new cluster candidates could
likely be the outcome once the survey be completed.

\acknowledgements
We thank David Nidever for providing us with an updated list of
SMASH fields. We are also grateful to the anonymous referee
for his/her suggestions to improve the Letter.

\bibliographystyle{aasjournal}

\begin{thebibliography}{}
\expandafter\ifx\csname natexlab\endcsname\relax\def\natexlab#1{#1}\fi

\bibitem[{{Belokurov} {et~al.}(2016){Belokurov}, {Erkal}, {Deason}, {Koposov},
  {De Angeli}, {Evans}, {Fraternali}, \& {Mackey}}]{belokurovetal2016}
{Belokurov}, V., {Erkal}, D., {Deason}, A.~J., {et~al.} 2016, ArXiv e-prints,
  arXiv:1611.04614

\bibitem[{{Bica} {et~al.}(2008){Bica}, {Bonatto}, {Dutra}, \&
  {Santos}}]{betal08}
{Bica}, E., {Bonatto}, C., {Dutra}, C.~M., \& {Santos}, J.~F.~C. 2008, \mnras,
  389, 678

\bibitem[{{Cioni} {et~al.}(2011){Cioni}, {Clementini}, {Girardi}, {Guandalini},
  {Gullieuszik}, {Miszalski}, {Moretti}, {Ripepi}, {Rubele}, {Bagheri},
  {Bekki}, {Cross}, {de Blok}, {de Grijs}, {Emerson}, {Evans}, {Gibson},
  {Gonzales-Solares}, {Groenewegen}, {Irwin}, {Ivanov}, {Lewis}, {Marconi},
  {Marquette}, {Mastropietro}, {Moore}, {Napiwotzki}, {Naylor}, {Oliveira},
  {Read}, {Sutorius}, {van Loon}, {Wilkinson}, \& {Wood}}]{cetal11}
{Cioni}, M.-R.~L., {Clementini}, G., {Girardi}, L., {et~al.} 2011, \aap, 527,
  A116

\bibitem[{{Contenta} {et~al.}(2016){Contenta}, {Gieles}, {Balbinot}, \&
  {Collins}}]{contentaetal2016}
{Contenta}, F., {Gieles}, M., {Balbinot}, E., \& {Collins}, M.~L.~M. 2016,
  ArXiv e-prints, arXiv:1611.06397

\bibitem[{{Dark Energy Survey Collaboration} {et~al.}(2016){Dark Energy Survey
  Collaboration}, {Abbott}, {Abdalla}, {Aleksi{\'c}}, {Allam}, {Amara},
  {Bacon}, {Balbinot}, {Banerji}, {Bechtol}, {Benoit-L{\'e}vy}, {Bernstein},
  {Bertin}, {Blazek}, {Bonnett}, {Bridle}, {Brooks}, {Brunner}, {Buckley-Geer},
  {Burke}, {Caminha}, {Capozzi}, {Carlsen}, {Carnero-Rosell}, {Carollo},
  {Carrasco-Kind}, {Carretero}, {Castander}, {Clerkin}, {Collett}, {Conselice},
  {Crocce}, {Cunha}, {D'Andrea}, {da Costa}, {Davis}, {Desai}, {Diehl},
  {Dietrich}, {Dodelson}, {Doel}, {Drlica-Wagner}, {Estrada}, {Etherington},
  {Evrard}, {Fabbri}, {Finley}, {Flaugher}, {Foley}, {Fosalba}, {Frieman},
  {Garc{\'{\i}}a-Bellido}, {Gaztanaga}, {Gerdes}, {Giannantonio}, {Goldstein},
  {Gruen}, {Gruendl}, {Guarnieri}, {Gutierrez}, {Hartley}, {Honscheid}, {Jain},
  {James}, {Jeltema}, {Jouvel}, {Kessler}, {King}, {Kirk}, {Kron}, {Kuehn},
  {Kuropatkin}, {Lahav}, {Li}, {Lima}, {Lin}, {Maia}, {Makler}, {Manera},
  {Maraston}, {Marshall}, {Martini}, {McMahon}, {Melchior}, {Merson}, {Miller},
  {Miquel}, {Mohr}, {Morice-Atkinson}, {Naidoo}, {Neilsen}, {Nichol}, {Nord},
  {Ogando}, {Ostrovski}, {Palmese}, {Papadopoulos}, {Peiris}, {Peoples},
  {Percival}, {Plazas}, {Reed}, {Refregier}, {Romer}, {Roodman}, {Ross},
  {Rozo}, {Rykoff}, {Sadeh}, {Sako}, {S{\'a}nchez}, {Sanchez}, {Santiago},
  {Scarpine}, {Schubnell}, {Sevilla-Noarbe}, {Sheldon}, {Smith}, {Smith},
  {Soares-Santos}, {Sobreira}, {Soumagnac}, {Suchyta}, {Sullivan}, {Swanson},
  {Tarle}, {Thaler}, {Thomas}, {Thomas}, {Tucker}, {Vieira}, {Vikram},
  {Walker}, {Wechsler}, {Weller}, {Wester}, {Whiteway}, {Wilcox}, {Yanny},
  {Zhang}, \& {Zuntz}}]{des2016}
{Dark Energy Survey Collaboration}, {Abbott}, T., {Abdalla}, F.~B., {et~al.}
  2016, \mnras, 460, 1270

\bibitem[{{Deason} {et~al.}(2016){Deason}, {Belokurov}, {Erkal}, {Koposov}, \&
  {Mackey}}]{deasonetal2016}
{Deason}, A.~J., {Belokurov}, V., {Erkal}, D., {Koposov}, S.~E., \& {Mackey},
  D. 2016, ArXiv e-prints, arXiv:1611.04600

\bibitem[{{Drlica-Wagner} {et~al.}(2016){Drlica-Wagner}, {Bechtol}, {Allam},
  {Tucker}, {Gruendl}, {Johnson}, {Walker}, {James}, {Nidever}, {Olsen},
  {Wechsler}, {Cioni}, {Conn}, {Kuehn}, {Li}, {Mao}, {Martin}, {Neilsen},
  {No{\"e}l}, {Pieres}, {Simon}, {Stringfellow}, {van der Marel}, \&
  {Yanny}}]{drlica-wagneretal2016}
{Drlica-Wagner}, A., {Bechtol}, K., {Allam}, S., {et~al.} 2016, ArXiv e-prints,
  arXiv:1609.02148

\bibitem[{{Feigelson} \& {Babu}(2012)}]{fb2012}
{Feigelson}, E.~D., \& {Babu}, G.~J. 2012, {Modern Statistical Methods for
  Astronomy}

\bibitem[{{Hammer} {et~al.}(2015){Hammer}, {Yang}, {Flores}, {Puech}, \&
  {Fouquet}}]{hammeretal2015}
{Hammer}, F., {Yang}, Y.~B., {Flores}, H., {Puech}, M., \& {Fouquet}, S. 2015,
  \apj, 813, 110

\bibitem[{{Kim} {et~al.}(2015){Kim}, {Jerjen}, {Mackey}, {Da Costa}, \&
  {Milone}}]{kimetal2015a}
{Kim}, D., {Jerjen}, H., {Mackey}, D., {Da Costa}, G.~S., \& {Milone}, A.~P.
  2015, \apjl, 804, L44

\bibitem[{{Kim} {et~al.}(2016){Kim}, {Jerjen}, {Mackey}, {Da Costa}, \&
  {Milone}}]{kimetal2016}
---. 2016, \apj, 820, 119

\bibitem[{{Mackey} {et~al.}(2016){Mackey}, {Koposov}, {Erkal}, {Belokurov}, {Da
  Costa}, \& {G{\'o}mez}}]{mackeyetal2016}
{Mackey}, A.~D., {Koposov}, S.~E., {Erkal}, D., {et~al.} 2016, \mnras, 459, 239

\bibitem[{{Martin} {et~al.}(2015){Martin}, {Nidever}, {Besla}, {Olsen},
  {Walker}, {Vivas}, {Gruendl}, {Kaleida}, {Mu{\~n}oz}, {Blum}, {Saha}, {Conn},
  {Bell}, {Chu}, {Cioni}, {de Boer}, {Gallart}, {Jin}, {Kunder}, {Majewski},
  {Martinez-Delgado}, {Monachesi}, {Monelli}, {Monteagudo}, {No{\"e}l},
  {Olszewski}, {Stringfellow}, {van der Marel}, \& {Zaritsky}}]{martinetal2015}
{Martin}, N.~F., {Nidever}, D.~L., {Besla}, G., {et~al.} 2015, \apjl, 804, L5

\bibitem[{{Martin} {et~al.}(2016){Martin}, {Jungbluth}, {Nidever}, {Bell},
  {Besla}, {Blum}, {Cioni}, {Conn}, {Kaleida}, {Gallart}, {Jin}, {Majewski},
  {Martinez-Delgado}, {Monachesi}, {Mu{\~n}oz}, {No{\"e}l}, {Olsen},
  {Stringfellow}, {van der Marel}, {Vivas}, {Walker}, \&
  {Zaritsky}}]{martinetal2016}
{Martin}, N.~F., {Jungbluth}, V., {Nidever}, D.~L., {et~al.} 2016, \apjl, 830,
  L10

\bibitem[{{Nidever} \& {Smash Team}(2015)}]{nideveretal2015}
{Nidever}, D., \& {Smash Team}. 2015, in Astronomical Society of the Pacific
  Conference Series, Vol. 491, Fifty Years of Wide Field Studies in the
  Southern Hemisphere: Resolved Stellar Populations of the Galactic Bulge and
  Magellanic Clouds, ed. S.~{Points} \& A.~{Kunder}, 325

\bibitem[{{Piatti}(2016)}]{piatti16}
{Piatti}, A.~E. 2016, \mnras, 459, L61

\bibitem[{{Piatti} {et~al.}(2016){Piatti}, {Ivanov}, {Rubele}, {Marconi},
  {Ripepi}, {Cioni}, {Oliveira}, \& {Bekki}}]{piattietal2016}
{Piatti}, A.~E., {Ivanov}, V.~D., {Rubele}, S., {et~al.} 2016, \mnras, 460, 383

\bibitem[{{Piatti} {et~al.}(2017){Piatti}, {et al.}}]{piattietal2017}
{Piatti}, A.~E., {et~al.} 2017, \prep

\bibitem[{{Pieres} {et~al.}(2016){Pieres}, {Santiago}, {Balbinot}, {Luque},
  {Queiroz}, {da Costa}, {Maia}, {Drlica-Wagner}, {Roodman}, {Abbott}, {Allam},
  {Benoit-L{\'e}vy}, {Bertin}, {Brooks}, {Buckley-Geer}, {Burke}, {Rosell},
  {Kind}, {Carretero}, {Cunha}, {Desai}, {Diehl}, {Eifler}, {Finley},
  {Flaugher}, {Fosalba}, {Frieman}, {Gerdes}, {Gruen}, {Gruendl}, {Gutierrez},
  {Honscheid}, {James}, {Kuehn}, {Kuropatkin}, {Lahav}, {Li}, {Marshall},
  {Martini}, {Miller}, {Miquel}, {Nichol}, {Nord}, {Ogando}, {Plazas}, {Romer},
  {Sanchez}, {Scarpine}, {Schubnell}, {Sevilla-Noarbe}, {Smith},
  {Soares-Santos}, {Sobreira}, {Suchyta}, {Swanson}, {Tarle}, {Thaler},
  {Thomas}, {Tucker}, \& {Walker}}]{pieresetal2016}
{Pieres}, A., {Santiago}, B., {Balbinot}, E., {et~al.} 2016, \mnras, 461, 519

\bibitem[{{Salem} {et~al.}(2015){Salem}, {Besla}, {Bryan}, {Putman}, {van der
  Marel}, \& {Tonnesen}}]{salemetal2015}
{Salem}, M., {Besla}, G., {Bryan}, G., {et~al.} 2015, \apj, 815, 77

\bibitem[{{Sales} {et~al.}(2016){Sales}, {Navarro}, {Kallivayalil}, \&
  {Frenk}}]{salesetal2016}
{Sales}, L.~V., {Navarro}, J.~F., {Kallivayalil}, N., \& {Frenk}, C.~S. 2016,
  ArXiv e-prints, arXiv:1605.03574

\bibitem[{{Stetson} {et~al.}(1990){Stetson}, {Davis}, \& {Crabtree}}]{setal90}
{Stetson}, P.~B., {Davis}, L.~E., \& {Crabtree}, D.~R. 1990, in Astronomical
  Society of the Pacific Conference Series, Vol.~8, CCDs in astronomy, ed.
  G.~H. {Jacoby}, 289--304

\bibitem[{{Valdes} {et~al.}(2014){Valdes}, {Gruendl}, \& {DES
  Project}}]{valdesetal2014}
{Valdes}, F., {Gruendl}, R., \& {DES Project}. 2014, in Astronomical Society of
  the Pacific Conference Series, Vol. 485, Astronomical Data Analysis Software
  and Systems XXIII, ed. N.~{Manset} \& P.~{Forshay}, 379

\bibitem[{{Vanderplas} {et~al.}(2012){Vanderplas}, {Connolly}, {Ivezi{\'c}}, \&
  {Gray}}]{astroml}
{Vanderplas}, J., {Connolly}, A., {Ivezi{\'c}}, {\v Z}., \& {Gray}, A. 2012, in
  Conference on Intelligent Data Understanding (CIDU), 47 --54

\end{thebibliography}

\appendix

\section{New MS stellar cluster candidates.}

\begin{figure}
\caption{2$\times$2 arcmin$^2$ $g$ images centred on the new
MS stellar cluster candidatess.  North is up and East to the left.}
\label{fig1A}
\end{figure}

\setcounter{figure}{2}
\begin{figure}
\caption{continued.}
\label{fig1A}
\end{figure}

\end{document}